\documentclass[aps,prd,twocolumn,superscriptaddress,floatfix]{revtex4-2}
\usepackage{comment}
\UseRawInputEncoding

\usepackage{graphicx}
\usepackage{booktabs}

\usepackage{amsmath,amssymb}

\usepackage{hyperref}
\usepackage[nameinlink,noabbrev]{cleveref}
\hypersetup{
  colorlinks=true,
  linkcolor=blue,
  urlcolor=blue,
  citecolor=blue
}

\usepackage{listings}
\usepackage{xcolor}
\usepackage{tikz}

\usepackage{xurl}
\usepackage{placeins}
\usepackage{indentfirst}

\usepackage{xparse}
\NewDocumentCommand{\figprelim}{O{} O{3.2} m}{%
\begin{tikzpicture}
  \node[inner sep=0] (img) {\includegraphics[#1]{#3}};
  \node[rotate=30, scale=#2, text opacity=0.20] at (img.center)
    {\bfseries PRELIMINARY};
\end{tikzpicture}%
}

\makeatletter
\newcommand{\chan}[1]{%
  \begingroup
  \edef\x{\endgroup\noexpand\nolinkurl{\detokenize{#1}}}%
  \x
}
\makeatother

\begin{document}

\title{
    Statistical Estimation and Correction of
    Model--Measurement Bias in Time--Dependent
    Correction Factors of KAGRA
}
\author{Shingo Hido} 
\email{shingo@icrr.u-tokyo.ac.jp}
\affiliation{Department of Physics, The University of Tokyo, 7-3-1 Hongo, Bunkyo-ku, Tokyo 113-0033, Japan}
\affiliation{Institute for Cosmic Ray Research, KAGRA Observatory, The University of Tokyo, 238 Higashi-Mozumi, Kamioka-cho, Hida City, Gifu 506-1205, Japan}

\author{Takahiro Yamamoto} 
\affiliation{Institute for Cosmic Ray Research, KAGRA Observatory, The University of Tokyo, 238 Higashi-Mozumi, Kamioka-cho, Hida City, Gifu 506-1205, Japan}

\author{Dan Chen} 
\affiliation{Gravitational Wave Science Project, National Astronomical Observatory of Japan, 2-21-1 Osawa, Mitaka, Tokyo 181-8588, Japan}

\author{Takahiro Sawada} 
\affiliation{Institute for Cosmic Ray Research, KAGRA Observatory, The University of Tokyo, 238 Higashi-Mozumi, Kamioka-cho, Hida City, Gifu 506-1205, Japan}

\author{Shinji Miyoki} 
\affiliation{Institute for Cosmic Ray Research, KAGRA Observatory, The University of Tokyo, 238 Higashi-Mozumi, Kamioka-cho, Hida City, Gifu 506-1205, Japan}

\date{\today}

\begin{abstract}

Calibration of gravitational-wave detectors reconstructs the strain
$h(t)$ from the detector output, and bias and uncertainty in this
reconstruction directly affect downstream analyses. In ground-based
interferometers, time-dependent correction factors (TDCFs) are estimated
from calibration lines to track temporal variations of the detector response, while the underlying model
parameters are periodically updated using broadband swept-sine calibration
measurements (SSCMs). However, if a model--measurement bias exists between
the measured transfer function and the reference model, the TDCFs inferred
from calibration lines can introduce a systematic deviation into the
reconstructed strain.
We propose a statistical framework to estimate and correct this bias using
repeated measurement-to-model ratios at the calibration-line frequencies.
The bias correction factors are estimated with a rolling random-effects
model based on restricted maximum likelihood (REML) and incorporated into
the TDCF estimation, with their uncertainty propagated to the reconstructed
response. Applying the method to KAGRA O4c data, we find that the
uncorrected response shows deviations of up to approximately 7\% in
magnitude and $5^\circ$ in phase relative to the SSCM-based reference in
representative examples. The correction reduces these deviations, with a
modest increase in the propagated uncertainty due to the included
correction-factor uncertainty. This framework provides a practical way to
combine broadband reference models with calibration-line-based tracking
when model--measurement bias is present.

\end{abstract}


\maketitle

\clearpage

\FloatBarrier
\newpage

\section{Introduction}

\subsection{Background}

The first direct detection of gravitational waves by Advanced Laser Interferometer Gravitational-wave Observatory (LIGO) \cite{Aasi_2015} marked the beginning of gravitational-wave astronomy as an observational science \cite{PhysRevLett.116.061102}. 
Since then, the global detector network has expanded into the LIGO--Virgo--KAGRA (LVK) collaboration \cite{Acernese_2015, 10.1093/ptep/ptaa125}, enabling increasingly sensitive and coherent observations of compact-binary coalescences.
The LVK transient catalogs have continued to grow through the fourth LIGO--Virgo--KAGRA Observing Run (O4), with the recently released GWTC-5.0 catalog reporting results through the second part of O4 and bringing the cumulative number of gravitational-wave candidates to 390 \cite{theligoscientificcollaboration2026gwtc50observationssecondfourth}.
O4 was carried out in multiple stages.
KAGRA participated in the first part of O4 and, after further commissioning and sensitivity improvements, rejoined the observing run during the third part of O4 (O4c), the KAGRA O4c calibration data analyzed in this work correspond to this later KAGRA observing period \cite{KAGRA_O4}.

As gravitational-wave detectors improve in sensitivity and the signal-to-noise ratio of observed events increases, calibration becomes increasingly important.
Calibration reconstructs the gravitational-wave strain h(t) from the detector output and therefore underpins essentially all downstream analyses, including low-latency searches, source-parameter estimation, and tests of general relativity.
Any bias or uncertainty in calibration propagates directly into the inferred waveform amplitude and phase, and can therefore affect astrophysical and fundamental-physics interpretations.
Previous studies have shown that calibration systematics can bias parameter estimation and potentially mimic apparent deviations from general relativity, with these issues becoming more pressing for future high-sensitivity detectors \cite{10.21468/SciPostPhysCommRep.5, s9jw-jg6n, Hall_2019}.

In ground-based interferometric detectors, the detector response is not perfectly stationary over long observing periods.
Time variations in optical gain, actuation strength, and related detector parameters must therefore be tracked during observing runs.
For this purpose, calibration lines, continuous excitations injected at selected frequencies, are widely used to estimate time-dependent correction factors (TDCFs) \cite{Tuyenbayev_2017}.
Because they rely on information at only a small number of discrete frequencies, TDCFs provide a practical framework for monitoring temporal changes in detector response and updating strain reconstruction without interrupting observations.
At the same time, an operational question remains: how should the information obtained from calibration lines be related to the broader swept-sine calibration measurements (SSCMs) that are performed periodically to estimate the parameters?
In particular, when a bias exists between the measurements and the reference model, the quantity tracked by calibration lines does not necessarily coincide with the quantity inferred from the SSCMs. 
This motivates the statistical problem addressed in this work.

\begin{figure}
\centering
\includegraphics[width=0.99\columnwidth]{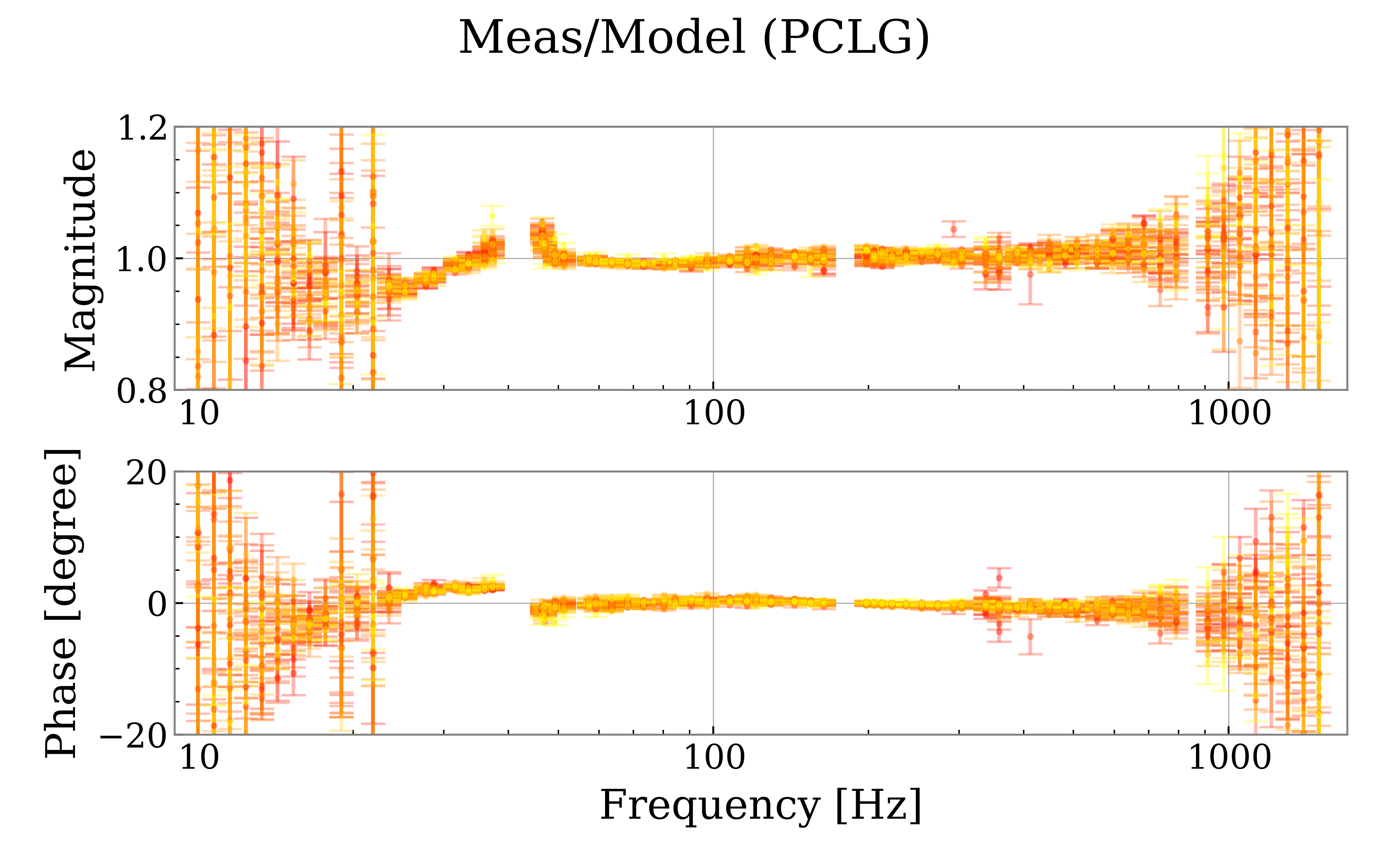}
\includegraphics[width=0.99\columnwidth]{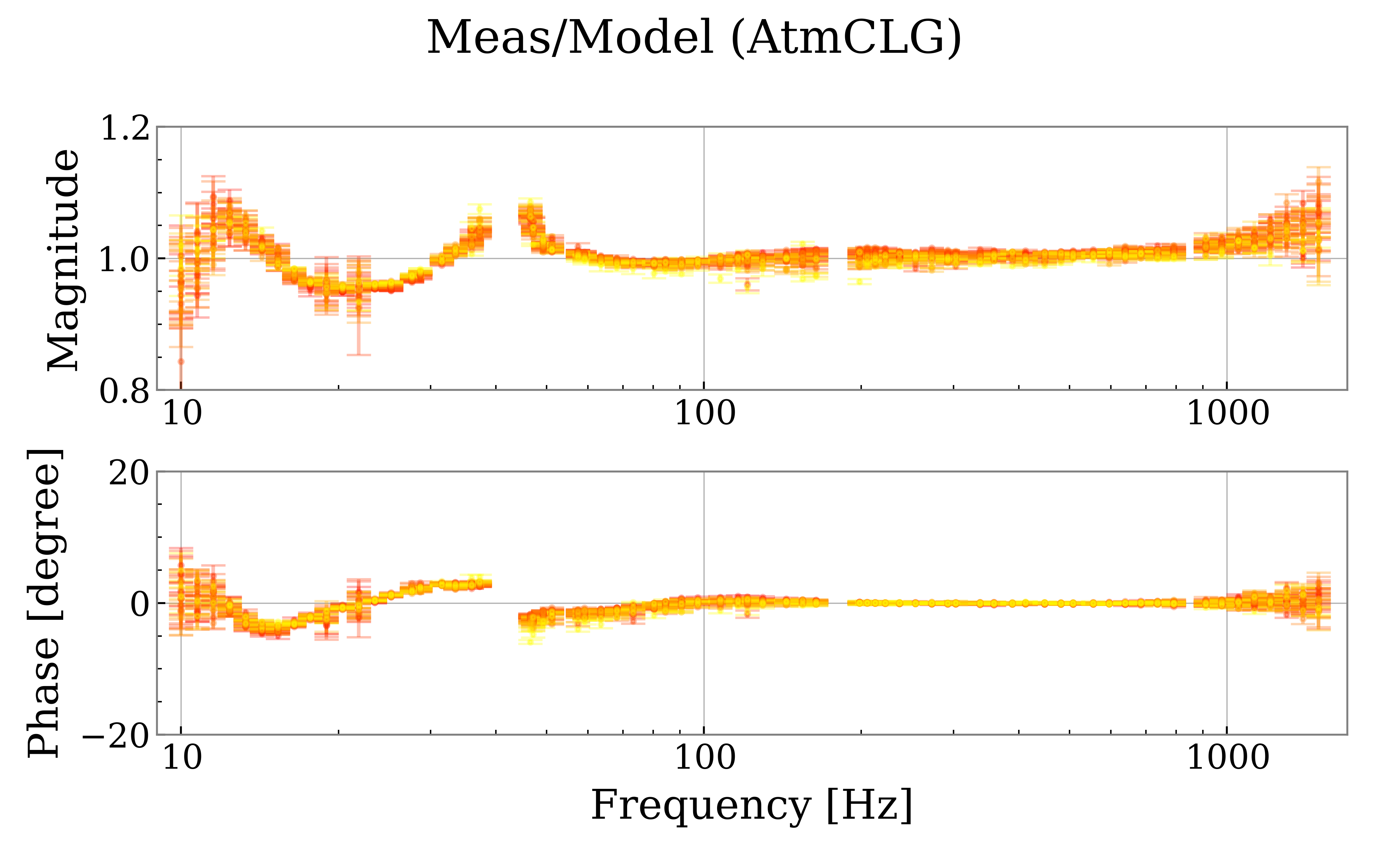}
\caption{\label{fig:tf_all}
Ratio of the measured transfer function to the best-fit model
(Meas/Model) for a two-month period starting in June 2025.
The upper group shows PCLG, the transfer function from the
photon-calibrator excitation to the DARM error signal, and the
lower group shows AtmCLG, the transfer function from the
test-mass-stage excitation to the DARM error signal.
In each group, the upper and lower panels show amplitude and
phase, respectively. The color indicates time, changing from red
in early June to yellow at later times. If the model fully described
the measurements, the ratios would be distributed around unity
in amplitude and around 0 degrees in phase; the observed
deviations indicate frequency-dependent model--measurement bias.
}
\end{figure}

\subsection{Problem setting and scope}
\label{sec:scope}

In KAGRA O4c, the calibration model parameters were regularly updated using SSCMs, while temporal variations during the observation run were tracked using calibration lines and the corresponding TDCFs. This calibration scheme is practical, but it can become problematic when the broadband measurements and the model are not fully consistent.

Fig.~\ref{fig:tf_all} shows two representative examples, obtained during KAGRA O4c, of the ratio of the measured transfer function to the best-fit model, hereafter referred to as Meas/Model.
The two quantities considered here are PCLG, the transfer function from the photon-calibrator \cite{Estevez_2021, Chen_2025} excitation to the differential arm (DARM) error signal, and AtmCLG, the transfer function from the test-mass actuator excitation to the DARM error signal.
Their precise definitions are given in Sec.~\ref{sec:data}. In the ideal case, this ratio would be distributed around unity in amplitude and around 0 degrees in phase. In practice, however, systematic deviations from these ideal values are observed. In this paper, we refer to such deviations as model-measurement bias.

\begin{figure}
\centering
\includegraphics[width=0.48\textwidth]{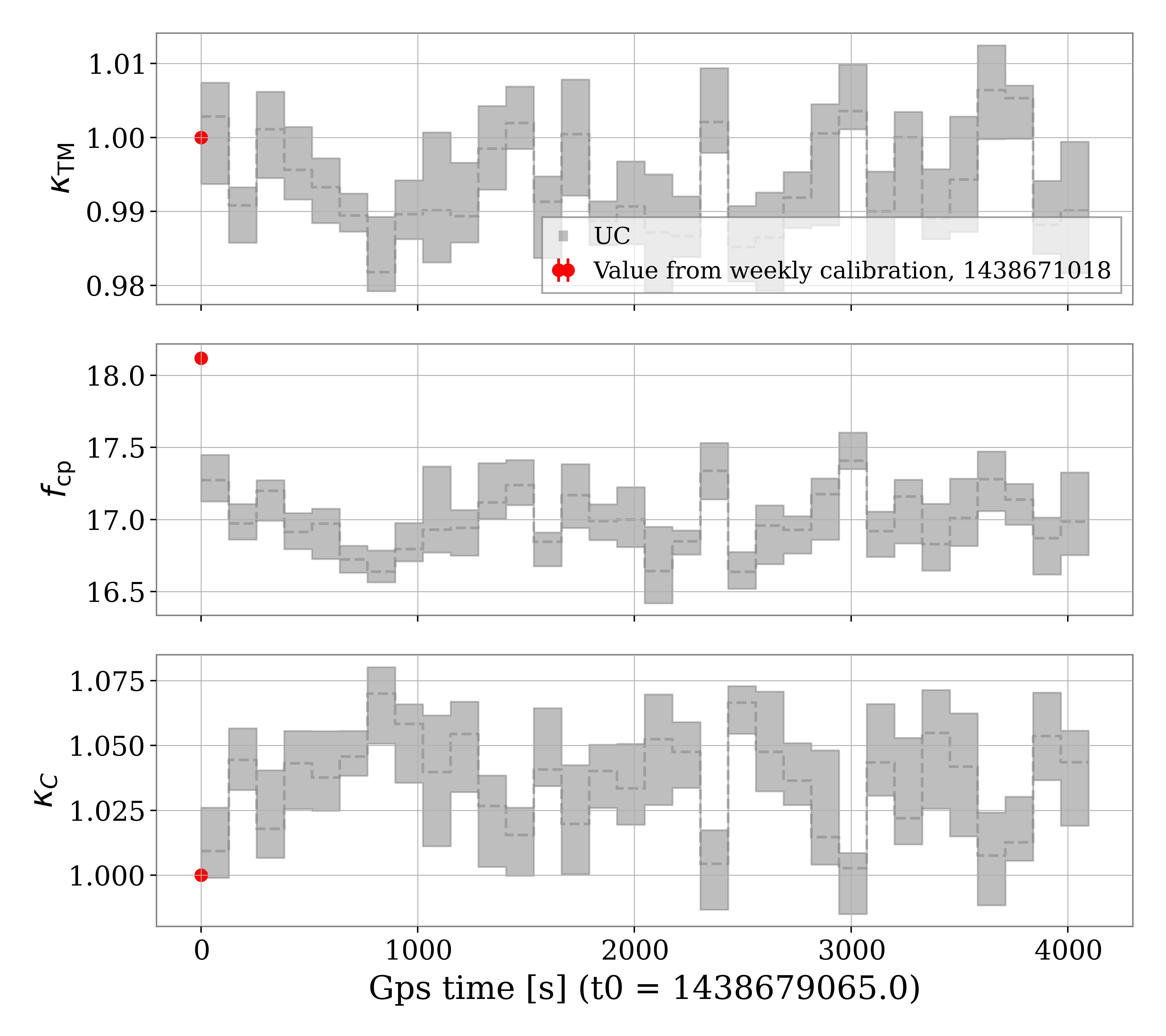}
\caption{\label{fig:tdcfs_original}
     Time series of the time-dependent correction factors (TDCFs): (top) relative test-mass-stage actuation efficiency, (middle) cavity pole, and (bottom) relative optical gain. The grey curves show the TDCFs estimated from calibration lines together with their uncertainty intervals. The label ``UC'' denotes the uncorrected case, in which no correction is applied for the model--measurement bias. Red markers show parameter estimates inferred from swept-sine calibration measurements (SSCMs) performed immediately before entering observing mode. The discrepancy between the grey curves and the red markers cannot be explained solely by the estimated TDCF uncertainty interval, suggesting a systematic offset. The GPS time in each label corresponds to the reference timestamp recorded prior to the measurement and therefore does not indicate the exact time at which the parameter estimate was obtained.
 }
\end{figure}

The practical consequence of this bias is illustrated in Fig.~\ref{fig:tdcfs_original}. The TDCFs inferred from the calibration lines are compared with the parameter estimates obtained from the SSCM performed immediately before the observing segment.
The observed discrepancy cannot be explained solely by the estimated TDCF uncertainty, indicating the presence of a systematic bias. If such biased TDCFs are directly applied in the reconstruction of $h(t)$, they can introduce an unnecessary bias into the calibrated strain.

The aim of this paper is to establish a statistical correction method for TDCF estimation in the presence of model-measurement bias.
Specifically, we estimate the model-measurement bias, derive a bias correction factor from it, apply this correction to the TDCFs, and propagate the associated uncertainty in a statistically consistent manner.
Using KAGRA O4c data, we demonstrate that this method provides a practical framework for correcting TDCF estimates and quantifying their uncertainty when model-measurement bias is present.

\section{Calibration framework and TDCF estimation}
\label{sec:data}

\begin{figure}
 \centering
    \includegraphics[width=0.45\textwidth]{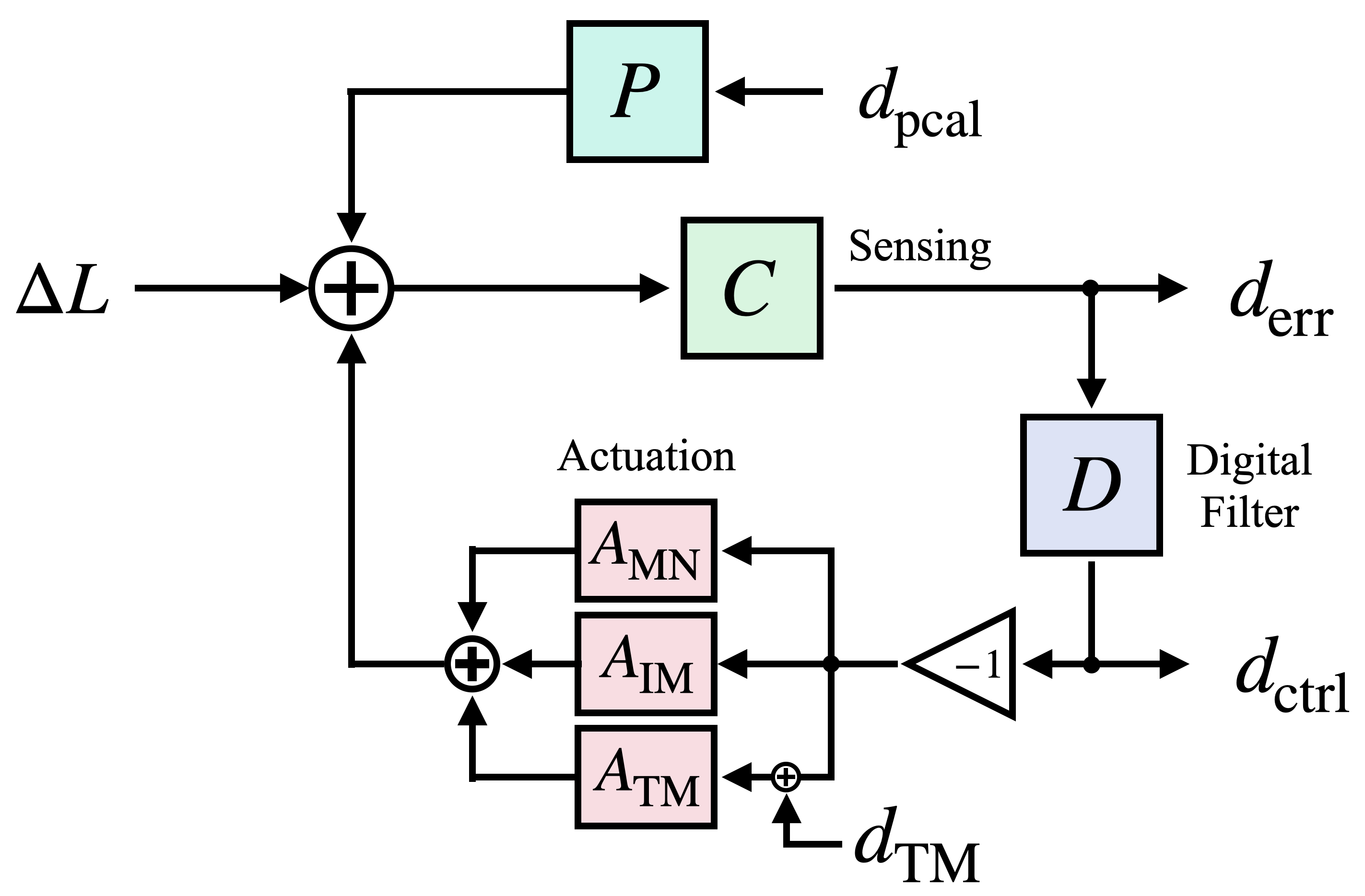}
 \caption{
    Schematic diagram of the DARM control loop. $P$ denotes the photon-calibrator function, $C$ the sensing function, 
    $D$ the digital filter, and $A_{\mathrm{TM}}$, $A_{\mathrm{IM}}$ and $A_{\mathrm{MN}}$ the actuation functions for 
    the test-mass (TM), intermediate-mass (IM), and marionette (MN) stages, respectively. 
    $\Delta L$ represents the DARM displacement; after subtraction of the feedback contribution, the residual signal passes through $C$ to form the error signal, $d_\mathrm{err}$, and through 
    $D$ to form the control signal, $d_\mathrm{ctrl}$. 
    $d_\mathrm{pcal}$ and $d_\mathrm{TM}$ indicate excitation points. Although not shown in the diagram, 
    equivalent excitation points also exist for the IM and MN stages.
 }
 \label{fig:darm}
\end{figure}

Fig.~\ref{fig:darm} shows a schematic diagram of the DARM control loop in KAGRA, which has a structure similar to that used in LIGO \cite{Sun_2020}.
The loop consists of the sensing function $C$, the digital filter $D$, and the actuation functions for the test-mass (TM), intermediate-mass (IM), and marionette (MN) stages,
denoted by $A_\mathrm{TM}$, $A_\mathrm{IM}$, $A_\mathrm{MN}$, respectively.
In this notation, the response function used to reconstruct the DARM displacement from the error signal $d_{\rm err}$ is written as
\begin{align}
    \label{eq:response}
    R(f, t) = \frac{1+C(f, t) D(f) A(f, t)}{C(f, t)}.
\end{align}
In this work, the sensing function is modeled as
\begin{equation}
    \label{eq:sen_model}
    C(f, t) = \frac{\kappa_C(t) H_C }{1 + if/f_{cp}(t)} F_C(f) e^{-2\pi i f \tau_C},
\end{equation}
and the actuation function is modeled as
\begin{align}
    \label{eq:act_model}
    A(f, t) &= A_{\mathrm{TM}}(f, t) + A_{\mathrm{IM}}(f, t) + A_{\mathrm{MN}}(f, t),
\end{align}
\begin{align}
    \label{eq:Ai_model}
    A_s(f, t) = \kappa_s(t)\, H_s\, F_s(f)\, e^{-2\pi i f \tau_A},
    \nonumber \\
    s \in \{\mathrm{TM}, \mathrm{IM}, \mathrm{MN}\}.
\end{align}
Here, \(H_C\) is the optical gain, \(\kappa_C(t)\) describes its relative temporal variation, \(f_{cp}(t)\) is the cavity pole frequency, \(\tau_C\) is the time delay in the sensing function, and \(F_C(f)\) represents the remaining frequency-dependent filters. 
Similarly, \(H_s\) is the actuation gain for stage \(s\), \(\kappa_s(t)\) describes its relative temporal variation, \(\tau_A\) is the time delay in the actuation function, and \(F_s(f)\) represents the remaining frequency-dependent filters in the actuation model~\cite{PhysRevD.96.102001, 10.1093/ptep/ptab018}.
The parameters \(\{H_C, f_{cp}, \tau_C, H_s, \tau_A\}\) are obtained from the SSCMs. In this paper, the TDCFs refer to the time-dependent parameters \(\kappa_C(t)\), \(f_{cp}(t)\), and \(\kappa_s(t)\) in this model.
Fig.~\ref{fig:darm} also shows the main excitation points used in this study.
$d_\mathrm{pcal}$ denotes the excitation applied through the photon calibrator path, modeled by $P$, and $d_\mathrm{TM}$ denotes the excitation injected into the test-mass actuation path. 
These excitation points are used in both the SSCMs and the calibration-line injections.
Following the notation introduced in Sec.~\ref{sec:scope}, PCLG denotes the transfer function from $d_\mathrm{pcal}$ to $d_\mathrm{err}$, $\mathrm{PCLG} = d_\mathrm{err}/d_\mathrm{pcal}$ and AtmCLG denotes the transfer function from $d_\mathrm{TM}$ to $d_\mathrm{err}$, $\mathrm{AtmCLG} = d_\mathrm{err}/d_\mathrm{TM}$.

\begin{table}[tb]
\centering
\caption{Calibration lines used for the TDCF estimation in O4c.}
\label{tab:cal_lines}
\begin{tabular}{lll}
\toprule
    Line & Frequency [Hz] & Excitation point \\
\midrule
    $f_\mathrm{TM}$ & $27.65$  & $d_{\mathrm{TM}}$ \\
    $f_\mathrm{pcal1}$ & $28.67$  & $d_{\mathrm{pcal}}$ \\
    $f_\mathrm{pcal2}$ & $31.53$  & $d_{\mathrm{pcal}}$ \\
\bottomrule
\end{tabular}
\end{table}

The calibration lines used for the TDCF estimation in O4c are summarized in Table~\ref{tab:cal_lines}.
No excitations were injected into the IM or MN stages.
Their frequencies are also included in the set of measurement frequencies used in the SSCMs.
The basic procedure for calculating the TDCFs follows the method described in \cite{Tuyenbayev_2017}.
To extract the complex amplitude information from the calibration lines, we used demodulation.
For each calibration-line frequency, we first applied a band-pass FIR filter with a passband of $\pm 0.8$ Hz centered on the target frequency, then demodulated the filtered signal at that frequency, and subsequently applied a low-pass FIR filter with a cutoff frequency of $0.1$ Hz.
This procedure yielded a complex amplitude time series associated with each calibration line.

\section{Statistical method for bias estimation, correction, and uncertainty propagation}
\subsection{REML-based estimation of the bias correction factor}

In this section, we describe how to define a bias correction factor from the Meas/Model values obtained from SSCMs.
The purpose is to define a correction factor that compensates for the model--measurement bias at the calibration-line frequencies, allowing the calibration-line information to be interpreted as the temporal variation of the parameters defined by the broadband SSCM fit.

For each SSCM set $i$, we consider the Meas/Model value at a calibration-line frequency $f_l$,
\begin{equation}
    \label{eq:mpm}
    y_i (f_l) = \frac{H_i^\mathrm{meas} (f_l)}{H_i^\mathrm{model} (f_l)}
\end{equation}
where $i$ labels the measurement set. 
In the following, the amplitude and phase of the complex ratio are treated separately.

The simplest choice for the bias correction factor is to use the Meas/Model value from the most recent SSCM set directly.
Although this approach reflects the latest measurement, it is sensitive to the statistical fluctuation of a single measurement and to occasional outliers.
On the other hand, a conventional weighted average based only on the quoted measurement uncertainties can underestimate the uncertainty when the scatter among measurement sets is larger than expected from those uncertainties alone.
Therefore, in order to define a representative bias correction factor, it is necessary to account not only for the uncertainty of each measurement point but also for the additional scatter between measurement sets.
This motivates the use of a random-effects model estimated with restricted maximum likelihood (REML) \cite{repec:bla:jorssa:v:172:y:2009:i:1:p:137-159, doi:10.3102/10769986030003261}.

Specifically, for either the amplitude or the phase of the Meas/Model ratio, each observed value $y_i$ is assumed to follow
\begin{equation}
    \label{eq:each_obsvalue}
    y_i | \mu_i \sim \mathcal{N} (\mu_i, a_i^2),
\end{equation}
where \(a_i^2\) 
denotes the variance of the corresponding amplitude or phase value of the Meas/Model value for the i-th SSCM set, obtained by propagating the measurement uncertainty of the SSCM transfer function \cite{BENDAT1978405}, and $\mu_i$ is the true bias associated with that measurement set. The true biases are further assumed to be distributed around a common mean $\mu$ as

\begin{equation}
    \label{eq:tru_dist}
    \mu_i \sim \mathcal{N} (\mu, \tau^2),
\end{equation}
where $\tau^2$ denotes the between-measurement variance.
These assumptions lead to 
\begin{equation}
    \label{eq:true_all}
    y_i | \mu, \tau \sim \mathcal{N}(\mu, a_i^2 + \tau^2)
\end{equation}
Under this model, the between-measurement variance $\tau^2$ is estimated by REML, and the pooled estimate of the common mean is obtained with weights
\begin{equation}
    \label{eq:true_weight}
    w_i = \frac{1}{a^2_i + \hat{\tau}^2},
\end{equation}
so that
\begin{equation}
    \label{eq:correction_factor}
    \hat{\mu} = \frac{\sum_i w_i y_i}{\sum_i w_i}.
\end{equation}
Here, the pooled estimate means the value obtained by combining multiple Meas/Model values while accounting for both the quoted uncertainty of each measurement set and the additional between-measurement scatter.
When $\hat{\tau}^2$ is close to zero, the scatter among measurement sets is largely explained by the quoted uncertainties alone.
In contrast, a positive $\hat{\tau}^2$ indicates non-negligible heterogeneity among measurement sets.
The confidence interval of $\tau^2$ is evaluated using the Q-profile method \cite{https://doi.org/10.1002/sim.2514}.

In this paper, we define the pooled estimate $\hat{\mu}$ at each calibration-line frequency as the bias correction factor.
The uncertainty associated with this estimate is characterized by the Hartung--Knapp-type standard error \cite{HK_HK2},
\begin{equation}
    \label{eq:se_hk}
    \mathrm{SE}_{\mathrm{HK}}(\hat{\mu})=\sqrt{q \, / \sum_i w_i},
\end{equation}
with
\begin{equation}
    \label{q}
    q=\frac{1}{k-1}\sum_i w_i (y_i-\hat{\mu})^2,
\end{equation}
where $k$ is the number of SSCM sets.
In practice, the estimate is updated in a rolling manner using the latest $k$ SSCM sets. 
In this paper, we adopt $k=5$ as the default choice and refer to this method as REML-R5. 
This is a pragmatic choice: previous simulation studies indicate that, under REML-based estimation, interval behavior is generally more reliable once at least several measurements are combined, whereas very small $k$ is more sensitive to small-sample effects \cite{K_num}. 
The robustness of this choice is assessed empirically in Sec.~\ref{sec:robust}.

\subsection{Bias-corrected TDCF estimation and uncertainty propagation}
\label{sec:corrtdcf}

The basic procedure for estimating the TDCFs from calibration lines follows Ref.~\cite{Tuyenbayev_2017, Viets_2018}.
For completeness, the explicit equations used to compute the TDCFs in this work are summarized in Appendix~\ref{sec:eq_tdcf}.
As summarized in Table~\ref{tab:cal_lines}, the calibration lines available in O4c enable the estimation of the three TDCFs, $f_{\mathrm{cp}}$, $\kappa_{\mathrm{C}}$, and $\kappa_{\mathrm{TM}}$.
Since no calibration lines are injected into the IM or MN stages, the corresponding factors are not treated in the present analysis.
The modification introduced in this work is that the complex transfer function obtained at each calibration-line frequency is corrected by the bias correction factor before being used in the TDCF equations.
Specifically, for a calibration line at frequency $f_l$, we replace the measured transfer function $H(t)\rvert_{f_l}$ by
\begin{equation}
    \label{eq:tf_corr}
    H^{\mathrm{corr}}(t)\rvert_{f_l}
    =
    \frac{H(t)\rvert_{f_l}}{\hat{\mu}(f_l)},
\end{equation}
where $\hat{\mu}(f_l)$ denotes the bias correction factor.
The corrected transfer functions are then substituted into the standard calibration-line equations to obtain bias-corrected estimates of the TDCFs.

To propagate the uncertainty of the bias correction factors to the TDCF estimates, we treat the correction factor at each calibration-line frequency as the value applicable until the next SSCM.
Accordingly, the relevant uncertainty is not limited to the uncertainty of the pooled estimate itself, but also includes the additional between-measurement variation associated with heterogeneity among SSCM sets.

This can be formulated as a hierarchical model. The pooled estimate at each calibration-line frequency is first represented by
\begin{equation}
    \label{eq:t_hk}
    \mu(f_l) \sim t_{k-1}\!\left(\hat{\mu}(f_l), \mathrm{SE}_{\mathrm{HK}}(\hat{\mu}(f_l))\right),
\end{equation}
where $t_{k-1}$ denotes a Student's $t$ distribution with $k-1$ degrees of freedom. The correction factor applied to the next interval is then modeled as
\begin{equation}
    \label{eq:n_tau}
    \theta_{\mathrm{next}}(f_l) \sim \mathcal{N}\!\left(\mu(f_l), \hat{\tau}^2(f_l)\right),
\end{equation}
where $\tau^2(f_l)$ is the REML estimate of the between-measurement variance. In this way, the uncertainty of the pooled estimate and the between-measurement variation are treated as distinct contributions.
In the main analysis, however, we adopt a simpler approximation by marginalizing this hierarchical model and representing the correction factor for the next interval by a single normal distribution,
\begin{equation}
    \label{eq:approx_n}
    \theta_{\mathrm{next}}(f_l) \approx \mathcal{N}\!\left(\hat{\mu}(f_l), \hat{\tau}^2(f_l) + \mathrm{SE}_{\mathrm{HK}}(\hat{\mu}(f_l))^2\right).
\end{equation}
In the parameter range relevant to this study, this approximation gives nearly the same uncertainty intervals as the explicit hierarchical construction while keeping the subsequent propagation procedure simple. A more detailed discussion of this approximation is given in Sec.~\ref{sec:robust}.

Using the distribution defined above, we propagate the uncertainty of the bias correction factors to the TDCF estimates by Monte Carlo sampling.
In the main analysis, the bias correction factors at different calibration-line frequencies, as well as their amplitude and phase components, are sampled independently.
A full treatment of these correlations would in principle be possible, but the number of available SSCM sets is limited, making their stable estimation difficult.
The impact of this simplification is examined in Sec.~\ref{sec:robust}.

For each draw of the bias correction factors, the corrected transfer function in Eq.~(\ref{eq:tf_corr}) is used and $\kappa_{\mathrm{TM}}$, $\kappa_{\mathrm{C}}$, and $f_{\mathrm{cp}}$ are recalculated from the corrected transfer functions.
Since TDCFs are typically estimated from data spanning a few minutes, we summarize them within each 128~s analysis segment by a representative value.
The representative value is defined as the segment median of the TDCF time series calculated using the representative bias correction factors $\hat{\mu}(f_l)$.

To evaluate the segment-wise uncertainty around this representative value, we generate $N=5000$ realizations.
For each realization, one joint moving-block bootstrap (MBB) resampling is applied to the three TDCF time series within each 128~s segment.
Specifically, overlapping non-circular 10~s blocks are sampled with replacement and concatenated until the original segment length is recovered, after which the median of each resampled TDCF series is taken as the segment-level statistic for that realization.
The MBB is applied jointly to the three TDCFs so that correlations among them are preserved within each realization.
The block length is set to 10~s based on the cutoff of the low-pass filter and the correlation times of the uncorrected TDCF time series.
The resulting ensemble of segment medians is used to estimate both the uncertainty interval and the correlation coefficients among the TDCFs for each segment.
The 16th and 84th percentiles of this ensemble are adopted as the uncertainty interval for each TDCF.
These correlation coefficients are then used to construct the segment-wise covariance matrix of the TDCFs for the multivariate normal sampling described in the next subsection.
The choice of the number of Monte Carlo realizations is discussed in Appendix~\ref{sec:boot}.

For comparison, we also evaluate the uncorrected case (UC), in which no bias correction is applied, and the last-sample method (LS), in which the most recent SSCM set alone is used to define the bias correction factor, together with the REML-R5 method introduced above.

\subsection{Evaluation of the response function using propagated TDCF uncertainties}

To evaluate how the TDCFs obtained with each bias-correction method affect the reconstruction of the detector response, we compare the response function constructed from the TDCFs with the reference response function derived from the SSCMs. For each method, we construct the response function $R_{\mathrm{type}}(f,t)$, defined by Eq.~(\ref{eq:response}), using the TDCFs and evaluate the complex ratio
\begin{equation}
    \frac{R_{\mathrm{type}}(f,t)}{R_{\mathrm{ref}}(f,t')},
\end{equation}
where $R_{\mathrm{ref}}(f,t')$ 
denotes the SSCM-based reference response associated with that segment, and $t'$ is the reference timestamp of the corresponding SSCM.

To propagate the TDCF uncertainties to this ratio, we construct, for each segment, a multivariate normal distribution
\begin{equation}
    \mathcal{N}(\boldsymbol{q}_{\rm{rep}},\boldsymbol{\Sigma}),
\end{equation}
for the TDCFs. Here, $\boldsymbol{q}_{\rm{rep}}$ is the vector of representative TDCFs for the segment, and $\boldsymbol{\Sigma}$ is the covariance matrix constructed from the segment-wise uncertainties and the correlations among the TDCFs. Its diagonal elements are given by the variances $\sigma_i^2$ of the individual TDCF parameters, while the off-diagonal elements are defined as
\begin{equation}
    \Sigma_{ij} = \rho_{ij}\,\sigma_i \sigma_j \qquad (i \neq j),
\end{equation}
where $\rho_{ij}$ is the correlation coefficient between the $i$th and $j$th TDCF parameters in the segment.
For each TDCF parameter $i$, the standard deviation is conservatively defined as
\begin{equation}
    \sigma_i
    =
    \max\!\left(
    \left|q_{84,i}-q_{\mathrm{rep},i}\right|,
    \left|q_{\mathrm{rep},i}-q_{16,i}\right|
    \right),
\end{equation}
where $q_{\mathrm{rep},i}$ is the representative TDCF, and $q_{16,i}$ and $q_{84,i}$ are the 16th and 84th percentiles of the bootstrap distribution, respectively.
This definition is adopted so that the uncertainty is not underestimated even when the bootstrap distribution is asymmetric.

For each segment, samples are drawn from the corresponding multivariate normal distribution $\mathcal{N}(\boldsymbol{q}_{\rm{rep}},\boldsymbol{\Sigma})$, and the response function $R_{\mathrm{type}}(f,t)$ is recalculated for each sample.
The number of samples assigned to each segment is taken to be nearly the same for all segments, so that the total number of samples over the full analysis period exceeds 50,000.
Using the full set of samples from all segments, we evaluate the ratio $R_{\mathrm{type}}(f, t)/R_{\mathrm{ref}}(f,t')$ at each frequency from 10~Hz to 1000~Hz in steps of 1~Hz.
Thus, the propagated uncertainty is characterized from the combined distribution over the entire analysis period, rather than evaluated independently for each segment.
At each frequency, the 16th and 84th percentiles of the resulting distribution are adopted as the propagated uncertainty interval.
This allows us to compare the systematic deviation and uncertainty in the $h(t)$ reconstruction arising from the TDCFs obtained with each bias-correction method.


\section{Results}
\label{sec:results}
In this section, we first examine the model--measurement bias across the three calibration-line frequencies used in the analysis.
We then present the TDCF estimates and the corresponding reconstruction-level comparisons for two representative 4096-s observing segments. These segments were selected from relatively long science-mode periods immediately following SSCMs, so that the response derived from the corresponding SSCMs can be used as a reference while retaining sufficient observing time for segment-by-segment comparison.
Within each 4096-s interval, the TDCFs are estimated independently for each 128-s segment.

\subsection{Model--measurement bias across calibration-line frequencies}
\label{sec:bias}

\begin{figure*}
\centering
\includegraphics[width=0.99\textwidth]{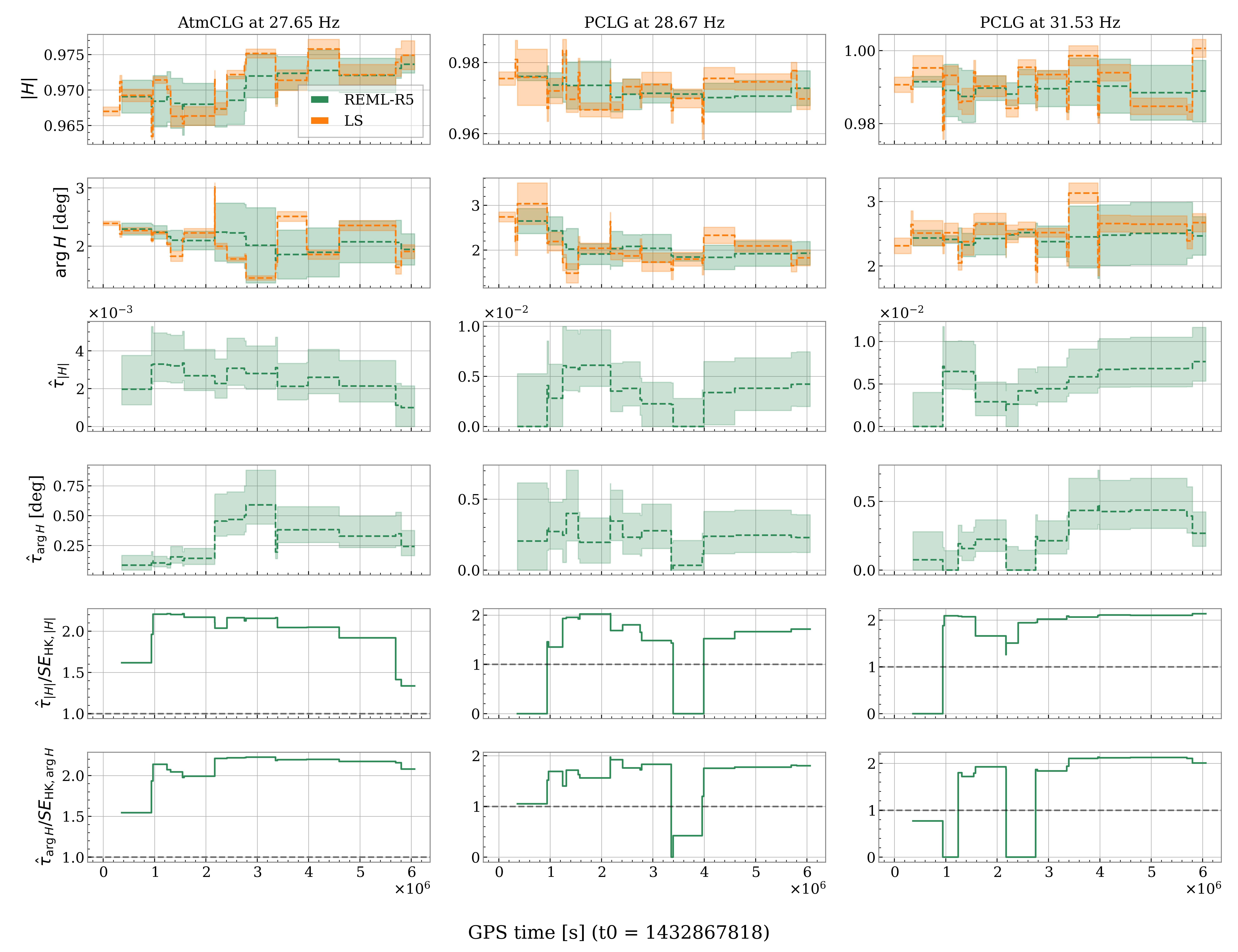}
\caption{\label{fig:mepmo_all}
    Model--measurement bias estimates across the three calibration-line frequencies used in the analysis. The columns correspond to AtmCLG at 27.65 Hz, PCLG at 28.67 Hz, and PCLG at 31.53 Hz.
The upper two rows show the bias estimates in amplitude and phase, the middle two rows show the estimated between-measurement heterogeneity $\hat{\tau}$ in amplitude and phase, and the lower two rows show the relative heterogeneity indicator, $\hat{\tau}/\mathrm{SE}_{\mathrm{HK}}$, in amplitude and phase.
For the bias estimates, LS (last sample) represents the latest value of Meas/Model, whereas REML-R5 represents the rolling random-effects restricted maximum likelihood (REML) estimate based on the latest five measurements.
For the heterogeneity panels, the solid line shows the estimated between-measurement heterogeneity $\hat{\tau}$, and the shaded bands indicate the 68\% confidence intervals.
The quantity $\hat{\tau}/\mathrm{SE}_{\mathrm{HK}}$ is shown as an indicator of the relative importance of between-measurement scatter in the uncertainty of the pooled bias estimate.
Across all three frequencies, the estimated heterogeneity is non-negligible over part of the observing period, indicating that the scatter among repeated SSCMs cannot always be explained solely by the quoted measurement uncertainties.}
\end{figure*}

Figure~\ref{fig:mepmo_all} summarizes the model--measurement bias estimates at the three
calibration-line frequencies used in the analysis: AtmCLG at 27.65~Hz,
PCLG at 28.67~Hz, and PCLG at 31.53~Hz. The figure also shows the
estimated between-measurement heterogeneity and the relative heterogeneity
indicator for the amplitude and phase components. These quantities allow us
to assess not only the central value of the bias correction factor, but also
whether the scatter among repeated SSCM results can be explained by the
quoted measurement uncertainties alone.

Across all three frequencies, the LS estimate directly follows the most recent
SSCM result, whereas the REML-R5 estimate provides a smoother correction
factor based on the latest five SSCM measurements. The uncertainty of
REML-R5 changes according to the scatter among these recent measurements:
it becomes larger when the repeated SSCM results are less consistent, and
smaller when they are mutually consistent. This behavior is expected because
REML-R5 accounts for both the quoted uncertainty of each SSCM result and
the additional between-measurement scatter.

The heterogeneity estimates show that the between-measurement scatter is appreciable 
over part of the observing period. The relative heterogeneity
indicator is also often non-negligible and sometimes exceeds unity, indicating
that this additional scatter can make an important contribution to the
uncertainty of the pooled bias estimate. Although the detailed time evolution
differs among the three calibration-line frequencies, the same qualitative
behavior is seen across the frequencies considered here. These observations
motivate the use of a bias-correction method that accounts for
between-measurement heterogeneity in the following TDCF and reconstruction
comparisons.

\subsection{TDCF estimates in representative segments}
\label{sec:tdcf}

\begin{figure*}
\centering
\includegraphics[width=0.48\textwidth]{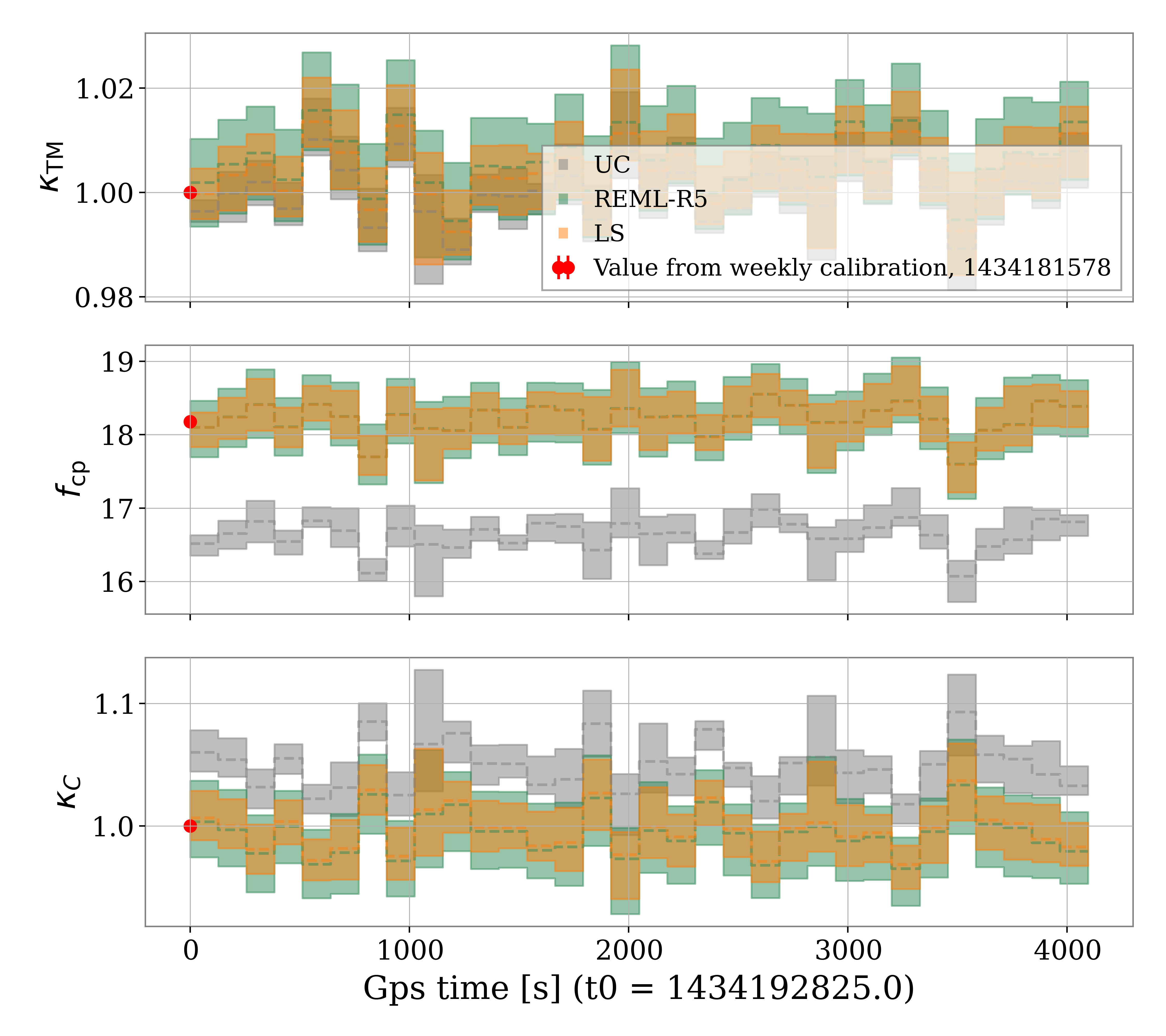}
\includegraphics[width=0.48\textwidth]{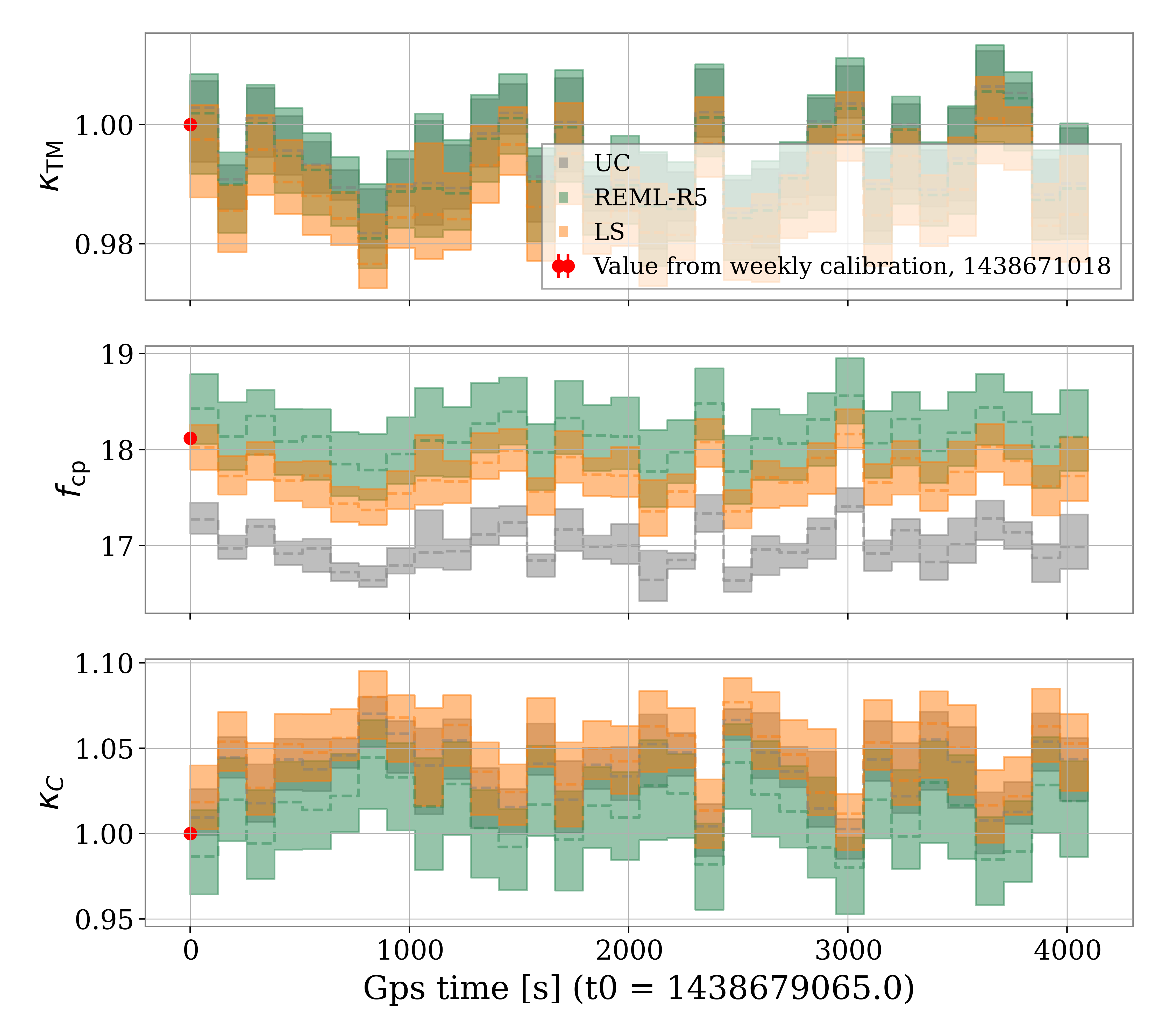}
\caption{\label{fig:tdcf_all}
Time series of the time-dependent correction factors (TDCFs) in two representative observing segments.
The left and right columns correspond to the segments referenced to GPS times 1434192825 and 1438679065, respectively.
In each column, the top, middle, and bottom panels show the relative test-mass-stage actuation efficiency $\kappa_{\mathrm{TM}}$, the cavity pole $f_{\mathrm{cp}}$, and the relative optical gain $\kappa_{\mathrm{C}}$, respectively.
The dashed curves show the segment median at each time, and the shaded bands indicate the 16th--84th percentiles of the ensemble of segment medians, estimated from the Monte Carlo realizations and joint moving-block bootstrap resampling.
The labels denote the bias-correction method: UC (uncorrected), LS (last sample), and REML-R5 (rolling random-effects restricted maximum likelihood estimate based on the latest five measurements).
Red markers show parameter estimates inferred from SSCMs performed immediately before entering observing mode.
In both segments, the bias-corrected results improve the consistency of the TDCF estimates with the SSCM-based reference relative to the uncorrected case.
The GPS time in each label corresponds to the reference timestamp recorded prior to the measurement and therefore does not indicate the exact time of the estimate.
    }
\end{figure*}

Figure~\ref{fig:tdcf_all} compares the TDCF estimates obtained with the UC, LS, and
REML-R5 methods in the two representative observing segments. The
SSCM-based parameter estimates obtained immediately before entering
observing mode are used as reference values.

In both segments, the UC estimate of $f_{\rm cp}$ clearly deviates from
the SSCM-based reference, consistent with the discrepancy already shown
in Fig.~\ref{fig:tdcfs_original}. After applying the bias correction, both LS and REML-R5 bring
$f_{\rm cp}$ into agreement with the SSCM-based reference within the
estimated 16th--84th percentile intervals. This demonstrates that the
bias correction removes the dominant TDCF-level deviation associated with
the model--measurement bias. The same qualitative tendency is also seen
for $\kappa_{\rm C}$, for which the bias-corrected
estimates are systematically shifted toward the SSCM-based reference
relative to the uncorrected case.

The difference between LS and REML-R5 is segment dependent. In the
segment referenced to GPS time 1438679065, REML-R5 gives more consistent
agreement with the SSCM-based reference across the three TDCFs, whereas
in the segment referenced to GPS time 1434192825, the LS and REML-R5
results are more similar. This behavior is consistent with the expectation
that the advantage of the pooled REML-R5 estimate becomes more visible
when the recent SSCM measurements exhibit stronger between-measurement
heterogeneity.

\subsection{Impact on strain reconstruction in representative segments}
\label{sec:sclg_r}

\begin{figure*}
\centering
\includegraphics[width=0.48\textwidth]{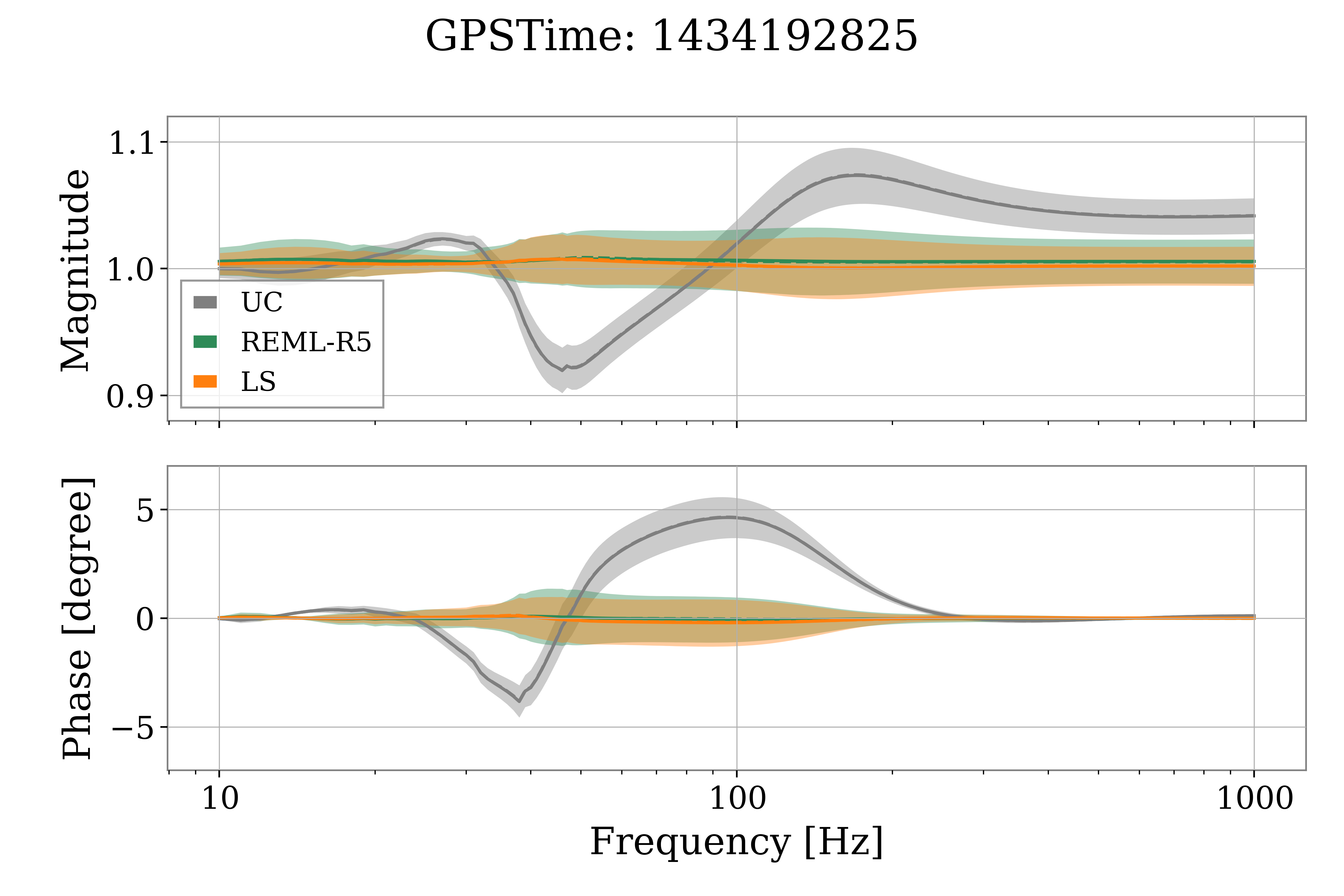}
\includegraphics[width=0.48\textwidth]{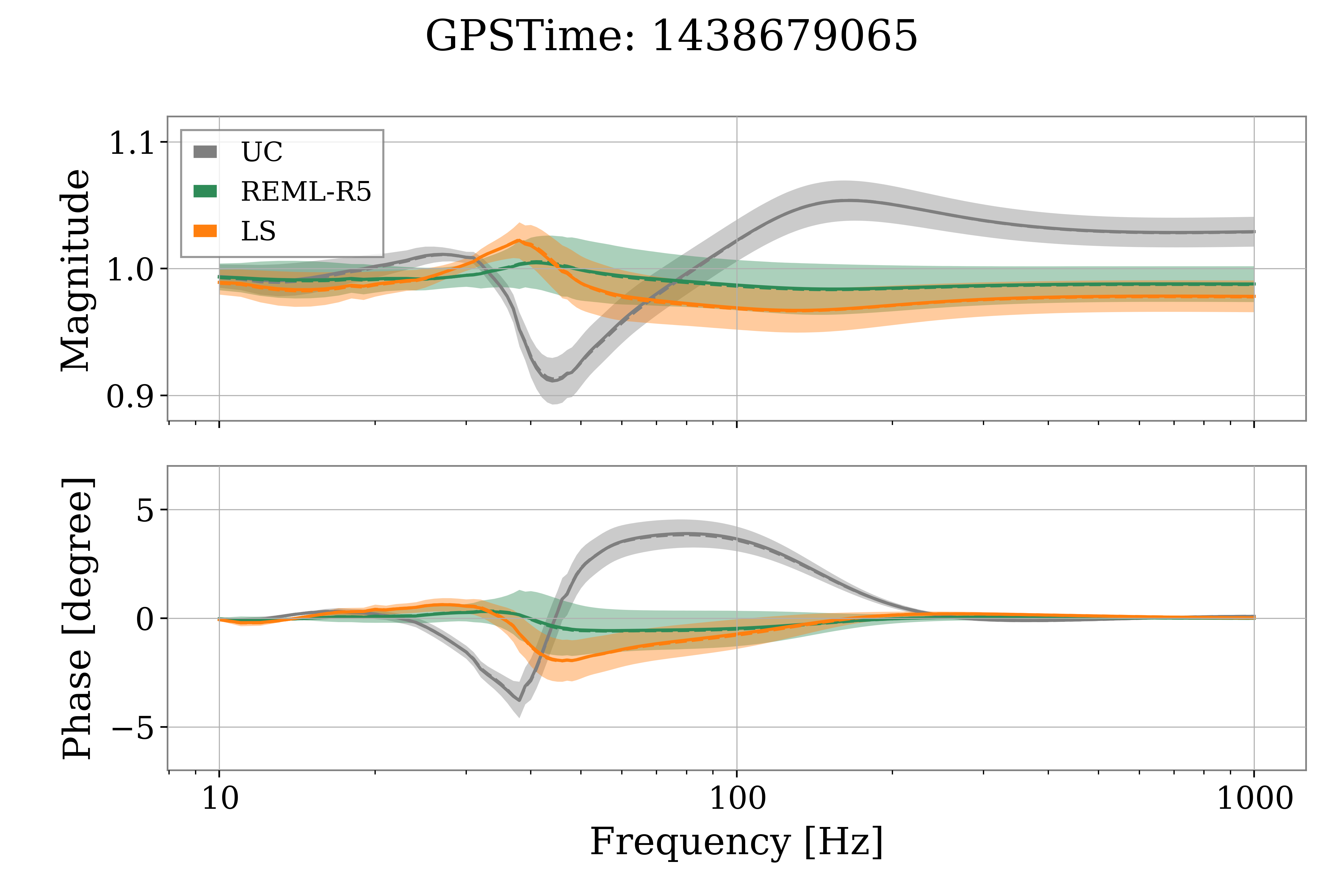}

\includegraphics[width=0.48\textwidth]{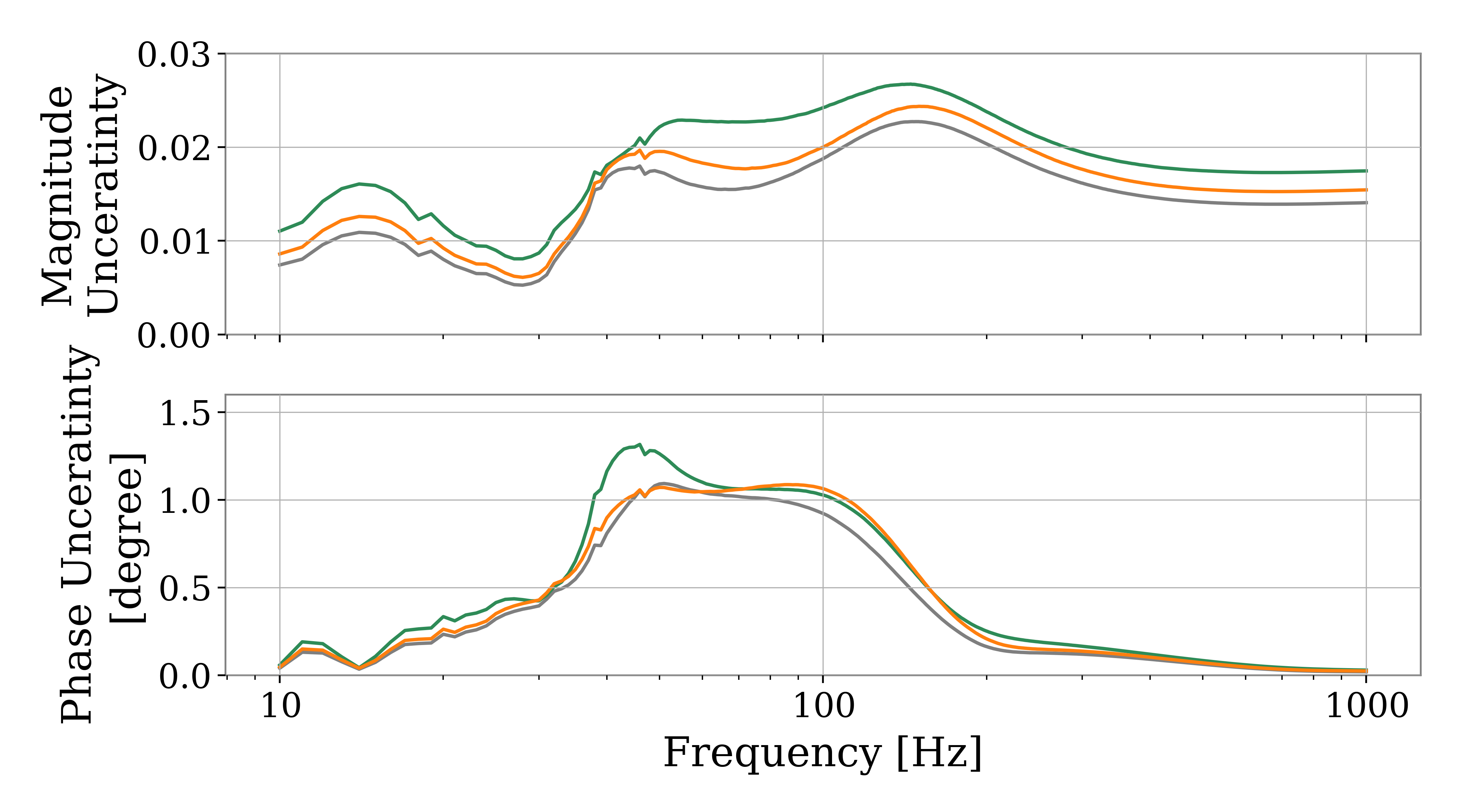}
\includegraphics[width=0.48\textwidth]{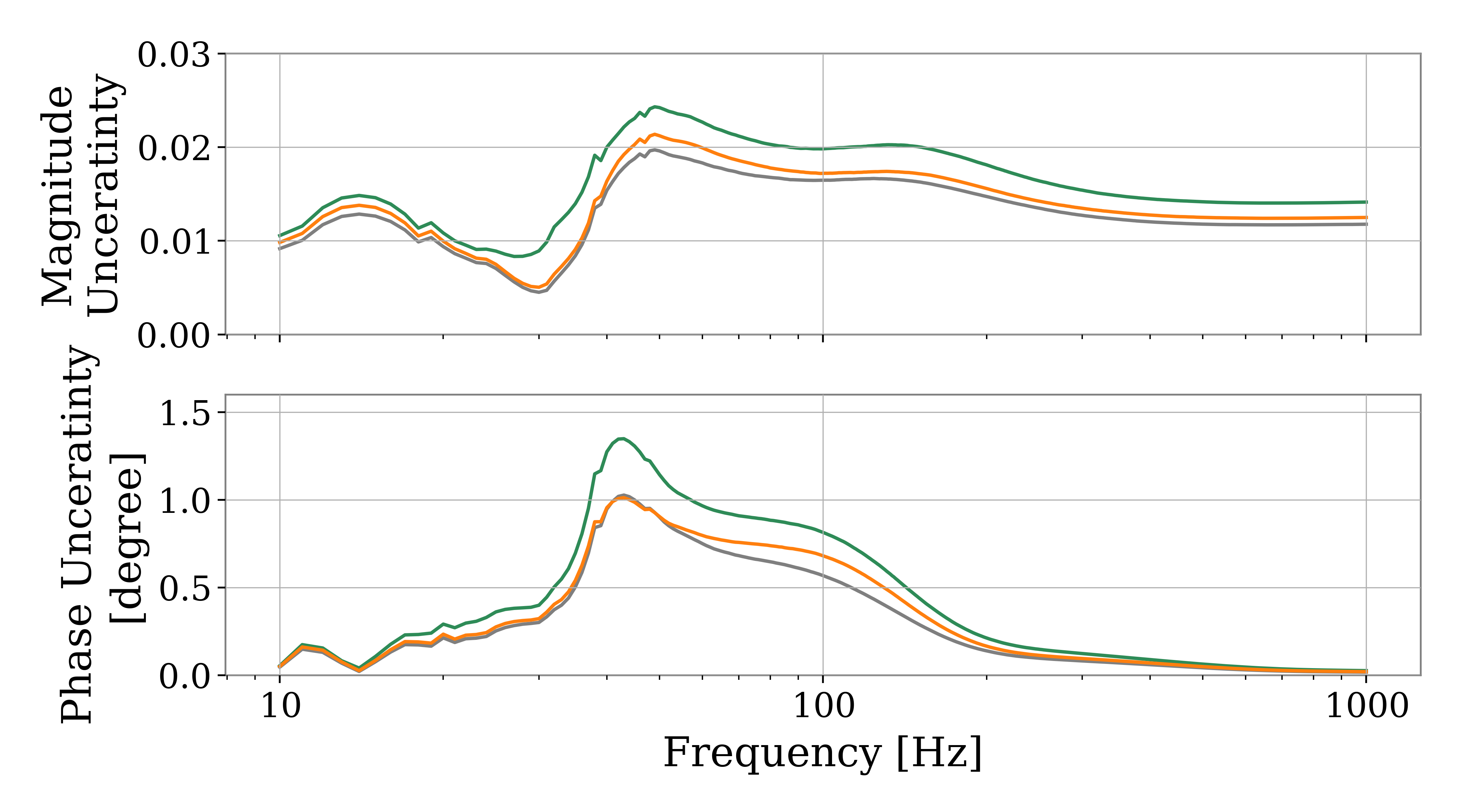}
\caption{\label{fig:sclg_all}
Frequency-domain comparison of $R_{\mathrm{type}}/R_{\mathrm{ref}}$ and its propagated uncertainty in two representative observing segments.
The left and right columns correspond to the segments referenced to GPS times 1434192825 and 1438679065, respectively.
In each column, the upper two panels show the magnitude and phase of $R_{\mathrm{type}}/R_{\mathrm{ref}}$, where $R_{\mathrm{type}}$ is the response obtained with each bias-correction method and $R_{\mathrm{ref}}$ is the reference response from the corresponding SSCMs.
The lower two panels show the half-widths of the 68\% propagated uncertainty intervals in magnitude and phase.
The labels denote the bias-correction method: UC (uncorrected), LS (last sample), and REML-R5 (rolling random-effects restricted maximum likelihood estimate based on the latest five measurements).
In both segments, applying the bias correction reduces the systematic deviation of $R_{\mathrm{type}}/R_{\mathrm{ref}}$ relative to the uncorrected case, while slightly increasing the propagated uncertainty interval.
}
\end{figure*}

Figure~\ref{fig:sclg_all} compares the reconstruction-level response ratios for the two
representative observing segments. For the UC case, clear systematic
deviations are seen in both magnitude and phase over part of the frequency
range in both segments, indicating that the model--measurement bias affects
the reconstructed strain. In these representative segments, the uncorrected
response deviates from the SSCM-based reference by up to approximately
$7\%$ in magnitude and $5^\circ$ in phase.

After applying the bias correction, these deviations are substantially reduced.
In the segment referenced to GPS time 1438679065, the difference between LS
and REML-R5 is more clearly visible, with REML-R5 remaining closer to unity
in magnitude and to zero in phase. In the segment referenced to GPS time
1434192825, the two bias-corrected results are more similar, although both
improve on the uncorrected case.

This reduction in systematic deviation is accompanied by a modest increase
in the propagated uncertainty interval. The broadening reflects the fact that,
once the bias correction factor is introduced explicitly, the uncertainty
associated with that correction is also incorporated into the uncertainty
budget. Thus, the bias correction reduces the response-level systematic
deviation while making the propagated uncertainty more complete.

\subsection{Robustness of the default choices}
\label{sec:robust}

\begin{figure}
\centering
\includegraphics[width=0.48\textwidth]{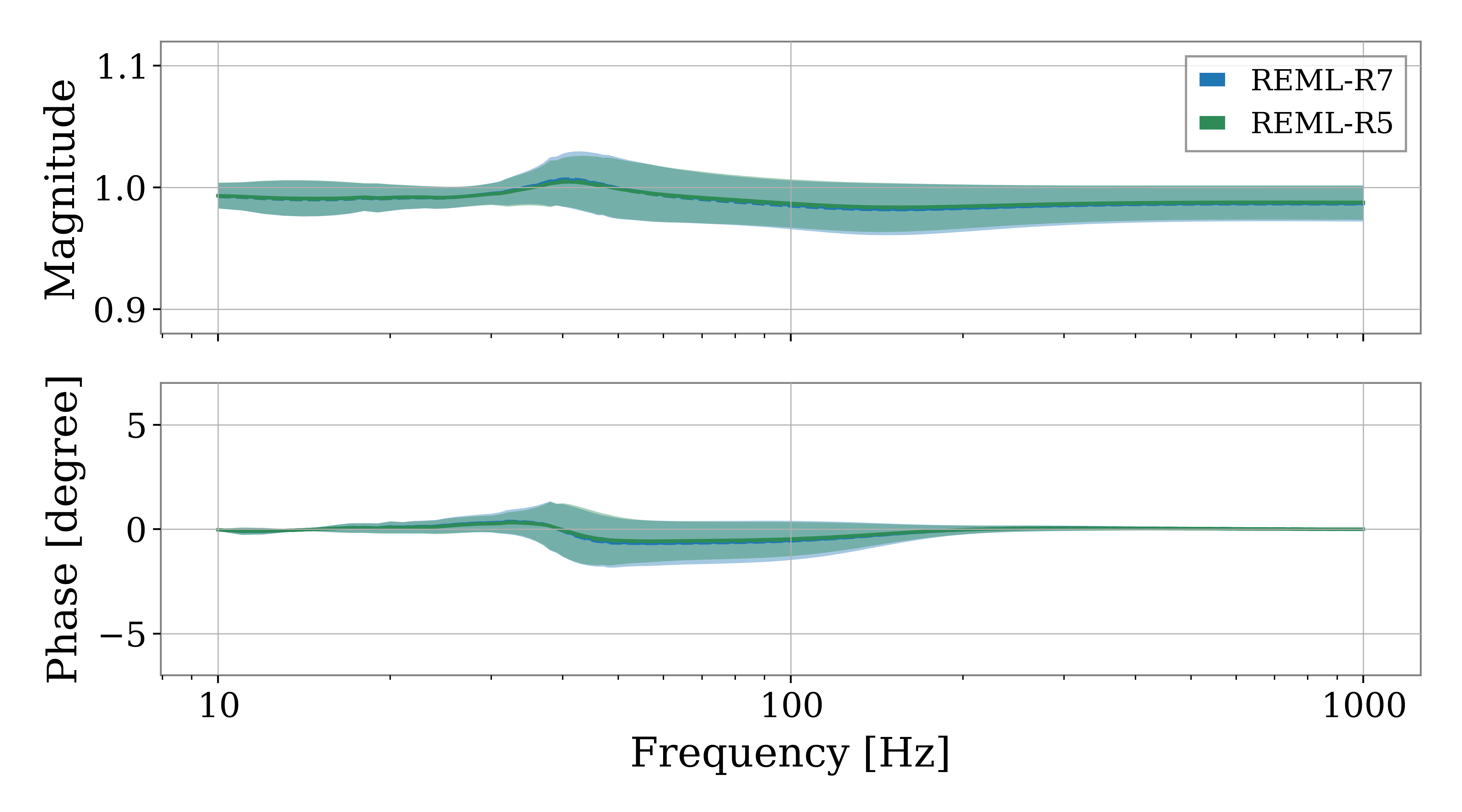}
\caption{\label{fig:RH}
Comparison of $R_{\mathrm{type}}/R_{\mathrm{ref}}$ obtained with rolling windows of five and seven measurements (R5 and R7).
The upper and lower panels show the magnitude and phase of $R_{\mathrm{type}}/R_{\mathrm{ref}}$, respectively.
The close agreement between R5 and R7 in these examples indicates limited sensitivity of the reconstruction-level result to whether five or seven measurements are used in the rolling window.
}
\end{figure}

\begin{figure}
\centering
\includegraphics[width=0.48\textwidth]{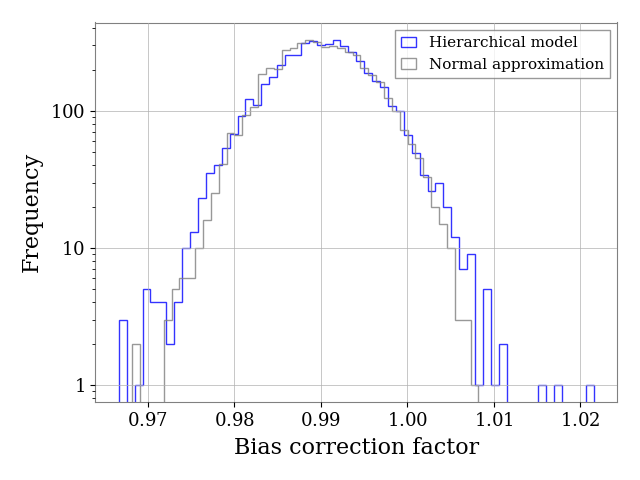}
\caption{\label{fig:hier}
    Comparison of the Monte Carlo sample distributions of the bias correction factor between the normal approximation and the hierarchical model. In the hierarchical model, the pooled mean is sampled from a $t$ distribution to reflect the small-sample uncertainty, while the between-measurement random effect is sampled from a normal distribution. 
    The two distributions agree closely in the central region, while small differences are visible mainly in the tails.
}
\end{figure}

\begin{figure}
\centering
\includegraphics[width=0.48\textwidth]{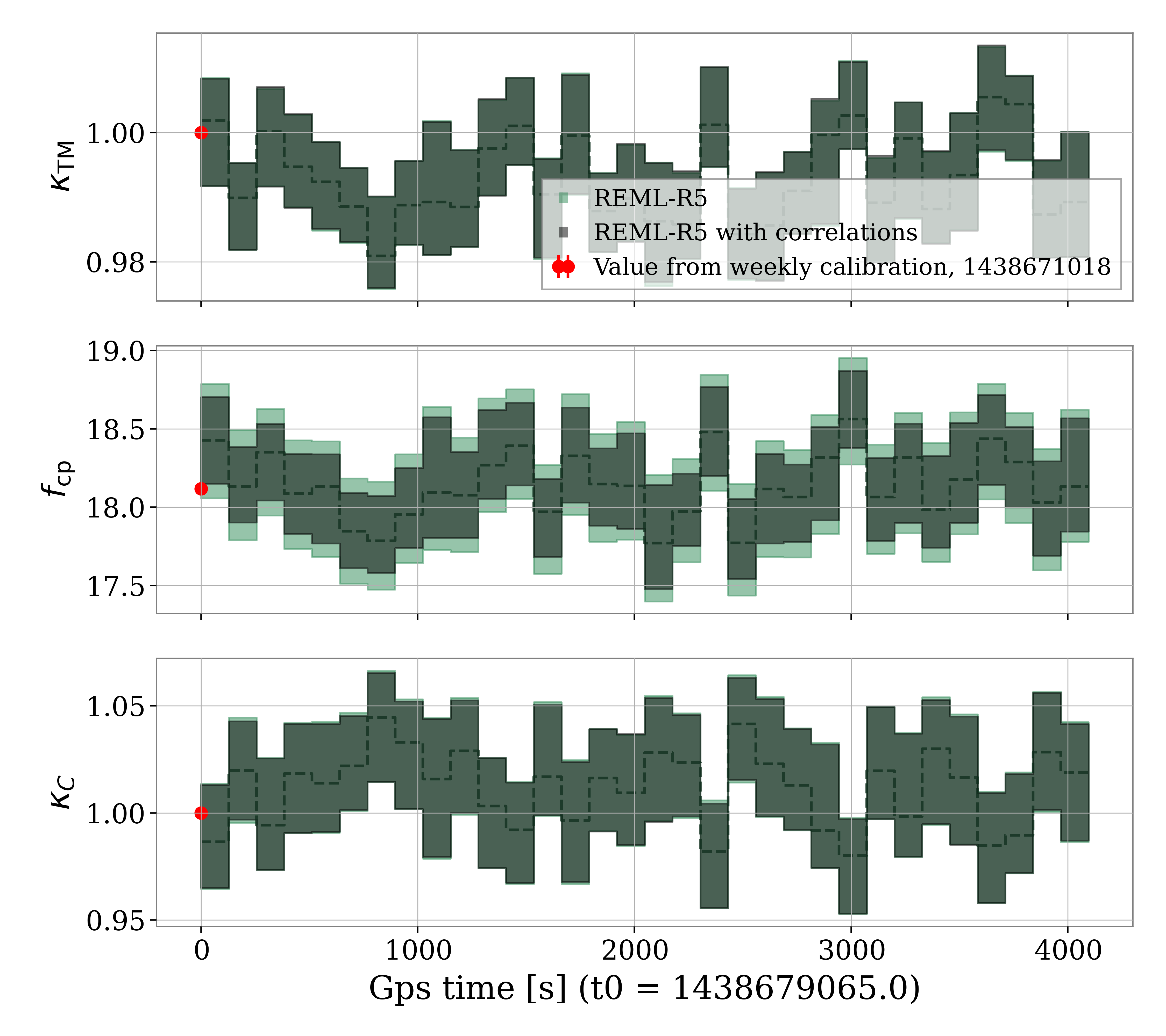}
\caption{\label{fig:corr}
    Comparison of the TDCF estimates obtained with the standard REML-R5 uncertainty propagation and with an additional diagnostic treatment of parameter correlations.
    In the latter case (``with correlations''), the largest pairwise correlations identified in the measured Meas/Model were conservatively set to unity and included in the joint sampling, while REML-R5 denotes the corresponding result without correlations.
    The upper, middle, and lower panels show $\kappa_{\mathrm{TM}}$, $f_{\mathrm{cp}}$, and $\kappa_{\mathrm{C}}$, respectively, for the representative GPS time 1438679065.
    Red markers show parameter estimates inferred from swept-sine calibration measurements performed immediately before entering observing mode.
    Only small differences are seen between the two cases; the main visible effect is a reduction in the uncertainty of $f_{\mathrm{cp}}$, while the other TDCFs remain nearly unchanged.
    This indicates that, for this example, the impact of parameter correlations on the propagated TDCF uncertainty is limited.
 }
\end{figure}

We briefly examine the robustness of the default choices adopted in the main analysis.
Figure~\ref{fig:RH} compares $R_{\mathrm{type}}/R_{\mathrm{ref}}$ obtained with rolling windows of five and seven measurements.
The close agreement between the two results indicates limited sensitivity of the reconstruction-level result to whether five or seven measurements are used in the rolling window.

Figure~\ref{fig:hier} compares the Monte Carlo sample distributions of the bias correction
factor obtained with the normal approximation and with the explicit hierarchical
model. The two distributions agree well overall, with the main difference
confined to the tails. In the example shown here, the 95\% interval is
$[0.9781, 1.0021]$ for the hierarchical model and $[0.9792, 1.0010]$ for the
normal approximation. The differences in the lower and upper endpoints are
both about $1.1\times10^{-3}$, corresponding to approximately $0.1\%$ in the
bias correction factor. This difference is sufficiently small for the present
uncertainty-propagation purpose, supporting the use of the simpler normal
approximation in the main analysis.

Figure~\ref{fig:corr} compares the TDCF estimates obtained with the standard REML-R5
uncertainty propagation and with a diagnostic treatment of parameter
correlations. 
Because the number of available SSCM sets is limited, we do not include the
full correlation structure of the bias correction factors in the default
analysis. As a diagnostic check, we estimated pairwise correlations among the
Meas/Model values, including the uncertainty of the correlation estimates, and
identified the pairs whose correlations were significantly different from zero.
For these pairs, we repeated the propagation after replacing the correlations
by their limiting values, $\rho=+1$ for positive correlations and $\rho=-1$ for
negative correlations.
Including these correlations mainly reduces the uncertainty of $f_{\rm cp}$, 
while the other TDCFs remain nearly unchanged. Therefore, neglecting these 
correlations in the default analysis does not underestimate the TDCF uncertainty 
in this example, and is conservative with respect to $f_{\rm cp}$.


\section{Discussion}

The results in Sec.~\ref{sec:results} show that introducing the bias correction
reduces the systematic deviation in the reconstructed response, while modestly
broadening the propagated uncertainty interval. Here we discuss how this behavior
should be interpreted, why the REML-R5-based treatment is a practical choice in
the present setting, and what limitations remain.

A central point is that the scatter among repeated SSCM results cannot always be
explained solely by their quoted measurement uncertainties. In the present data,
the relative heterogeneity indicator is often non-negligible and sometimes exceeds
unity, indicating that between-measurement scatter makes an important contribution
to the uncertainty of the pooled bias estimate. In such a situation, a correction
based only on the most recent SSCM may not adequately represent either the
representative correction value or its uncertainty for the period until the next
SSCM. The rolling REML estimate based on the most recent five measurements
provides a practical alternative, because it combines recent information while
accounting for between-measurement heterogeneity. This interpretation is broadly
consistent with the discussion of REML-based interval behavior in \cite{K_num}.
The use of five measurements should not be regarded as theoretically optimal.
Nevertheless, in the present data the heterogeneity is not negligible and the
individual SSCM uncertainties are not strongly imbalanced, making the REML-R5
pooled estimate a reasonable predictive quantity for the period between SSCMs.

Another important feature of the present results is that the reduction of
systematic deviation is accompanied by a modest increase in propagated
uncertainty. This does not indicate a degradation of the method. Rather, once the
bias correction factor is introduced explicitly, the uncertainty associated with
that correction must also be included in the uncertainty budget. In the present
examples, this contribution can be comparable to, or even larger than, the
uncertainty arising from the intrinsic time variation of the TDCFs themselves.
The wider propagated interval obtained after correction therefore more
realistically reflects the uncertainty relevant to the corrected reconstruction.

At the same time, the present implementation remains a first-order statistical
treatment. The REML estimate of $\hat{\tau}^2$ is treated as fixed during propagation,
and its own estimation uncertainty is not incorporated into the predictive
distribution. Correlations across calibration-line frequencies and between
amplitude and phase are also not fully incorporated in the main analysis, because
the number of available SSCM sets is limited for stable estimation of the full
correlation structure. In addition, the reconstruction-level validation is limited
to representative observing segments rather than a full-run validation.
Nevertheless, for $f_{\mathrm{cp}}$, which is the most influential parameter in
the present sensing correction, we additionally confirmed over the O4c observing
period that the proposed correction tends to reduce the bias relative to the
uncorrected case. Together with the representative reconstruction-level results
shown in Sec.~\ref{sec:results}, this supports the interpretation that the
reduction of the reconstruction bias is driven primarily by the improvement in
$f_{\mathrm{cp}}$.

Despite these limitations, the issue addressed here is not unique to KAGRA.
Whenever a calibration scheme combines a broadband reference model with
time-dependent correction factors inferred from calibration lines, residual
model--measurement inconsistency can propagate into the reconstructed response if
left untreated. In that sense, the framework developed here may also be useful in
other calibration settings where similar mismatches arise.


\section{Conclusion}

We have presented a statistical framework to estimate and correct
model--measurement bias in the time-dependent correction factors (TDCFs)
used for KAGRA calibration. In representative KAGRA O4c examples, the
uncorrected response deviated from the SSCM-based reference by up to
approximately 7\% in magnitude and 5$^\circ$ in phase, and these deviations
were reduced by the proposed bias correction. The propagated uncertainty
interval became slightly wider because the uncertainty of the bias correction
factor was explicitly included. These results support the proposed method for
calibration settings in which calibration-line tracking is combined with a
broadband reference model in the presence of residual model--measurement bias.

\FloatBarrier
\begin{acknowledgments}

This work was supported by MEXT, JSPS Leading-edge Research Infrastructure Program, JSPS Grant-in-Aid for Specially Promoted Research 26000005, JSPS Grant-in-Aid for Scientific Research on Innovative Areas 2402: 24103006, 24103005, and 2905: JP17H06358, JP17H06361 and JP17H06364, JSPS Core-to-Core Program A. Advanced Research Networks, JSPS Grant-in-Aid for Scientific Research (S) 17H06133 and 20H05639 , JSPS Grant-in-Aid for Transformative Research Areas (A) 20A203: JP20H05854, the joint research program of the Institute for Cosmic Ray Research, University of Tokyo, National Research Foundation (NRF), Computing Infrastructure Project of Global Science experimental Data hub Center (GSDC) at KISTI, Korea Astronomy and Space Science Institute (KASI), and Ministry of Science and ICT (MSIT) in Korea, Academia Sinica (AS), AS Grid Center (ASGC) and the National Science and Technology Council (NSTC) in Taiwan under grants including the Science Vanguard Research Program, Advanced Technology Center (ATC) of NAOJ, and Mechanical Engineering Center of KEK.

\end{acknowledgments}

\appendix

\section{Bootstrap settings}
\label{sec:boot}

\begin{figure}
 \centering
    \includegraphics[width=0.45\textwidth]{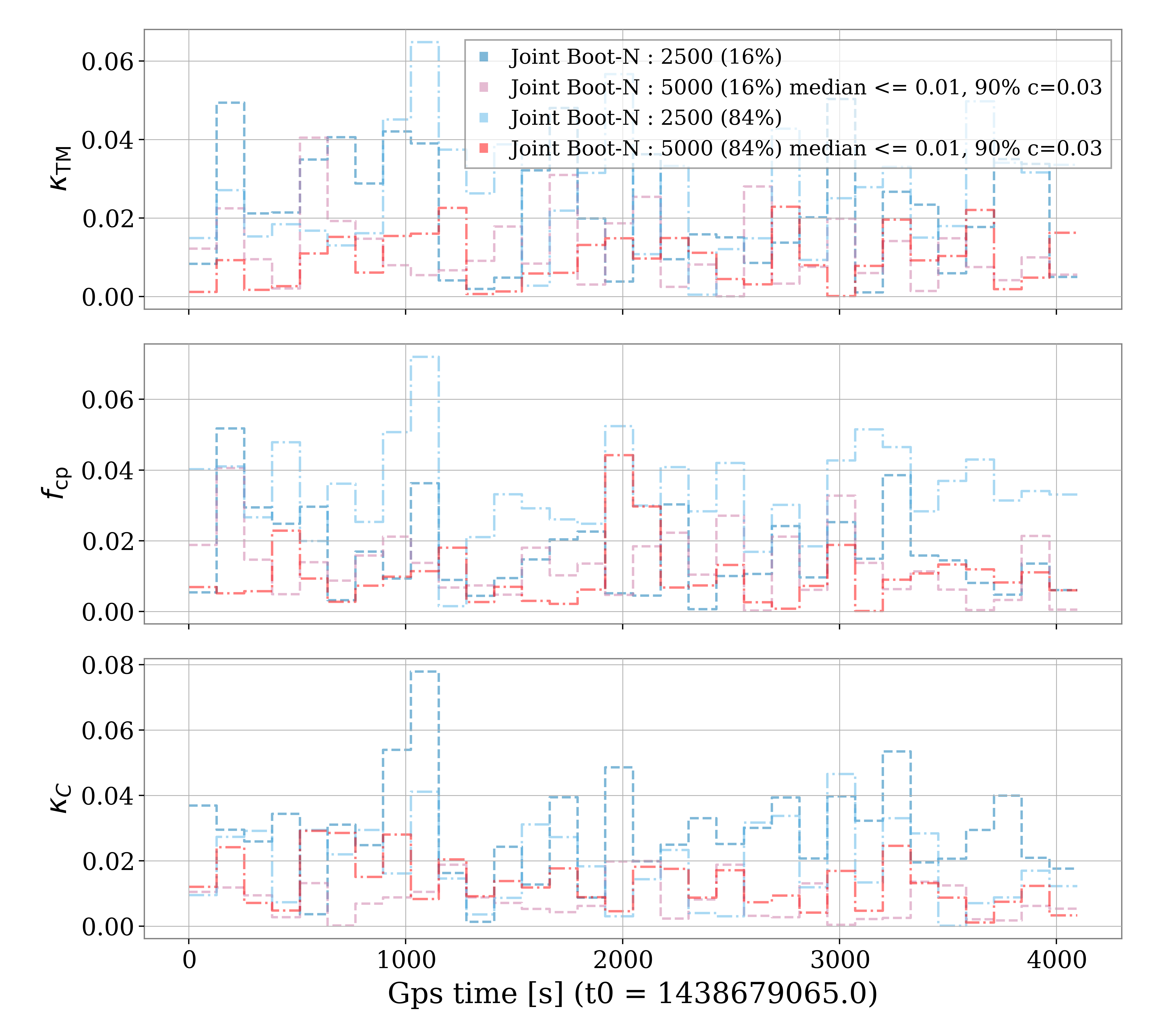}
 \caption{
     Stability of the estimated 68\% uncertainty intervals against the bootstrap sample number $N$.
    For each $N$, two independent runs with different random seeds are compared using
    $Q_p(t)=\lvert q_p^{(0)}(t)-q_p^{(1)}(t)\rvert / w(t)$ for $p=16\%$ and $84\%$,
    where $w(t)$ is the mean 68\% interval width of the two runs.
    The top, middle, and bottom panels show the results for $\kappa_{\mathrm{TM}}$, $f_{\mathrm{cp}}$, and $\kappa_C$, respectively.
    The legend gives the median and 90th percentile of $Q_p$.
    We adopt $N=5000$, for which the percentile estimates are sufficiently stable.
}
 \label{fig:tdcfs_joint_num}
\end{figure}

In the main analysis, the uncertainty interval of each TDCF estimate is defined by the 16th and 84th percentiles of the moving-block-bootstrap realizations, using $N=5000$ Monte Carlo samples for the bias-correction factors and the subsequent TDCF calculation. In this appendix, we briefly examine the stability of the estimated uncertainty intervals with respect to the sample number $N$.

Figure~\ref{fig:tdcfs_joint_num} shows the stability of the estimated 68\% uncertainty intervals as a function of $N$. For each $N$, two independent runs with different random seeds are compared through
\begin{equation}
    Q_p(t) = \frac{\left| q_p^{(0)}(t) - q_p^{(1)}(t) \right|}{w(t)},
\end{equation}
where $p=16\%$ or $84\%$, $q_p^{(0)}(t)$ and $q_p^{(1)}(t)$ are the corresponding percentile estimates from the two runs, and
\begin{equation}
\begin{aligned}
w(t)
= \frac{1}{2}\Big[
&\left(q_{84}^{(0)}(t)-q_{16}^{(0)}(t)\right) \\
&+ \left(q_{84}^{(1)}(t)-q_{16}^{(1)}(t)\right)
\Big]
\end{aligned}
\end{equation}
is the mean width of the two estimated 68\% intervals.

As $N$ increases, the discrepancy between the two runs decreases, indicating that the percentile estimates are becoming numerically stable. In the present analysis, $N=5000$ was adopted because it provides sufficiently stable uncertainty intervals across all analyzed chunks: the median of $Q_p$ is below 1\%, and its 90th percentile is below 3\%. This level of variation is small compared with the uncertainty interval itself and is therefore adequate for the purposes of the present uncertainty propagation.

We therefore conclude that the default choice $N=5000$ is sufficient to obtain stable bootstrap-based uncertainty estimates for the TDCFs.
\section{Equations for TDCF estimation}
\label{sec:eq_tdcf}

In this appendix, we summarize the explicit equations used to compute the TDCFs discussed in Sec.~\ref{sec:corrtdcf}.
The test-mass actuation correction factor $\kappa_\mathrm{TM}$ is computed as
\begin{equation}
\label{eq:kappa_tm}
\begin{aligned}
\kappa_\mathrm{TM}
&=
\left.\frac{d_\mathrm{err}}{d_\mathrm{TM}}\right|_{f_\mathrm{TM}}
\left.\frac{d_\mathrm{pcal}}{d_\mathrm{err}}\right|_{f_\mathrm{pcal1}}
\frac{
    H^\mathrm{model}_\mathrm{PCLG}(f_\mathrm{pcal1})
}{
    H^\mathrm{model}_\mathrm{PCLG}(f_\mathrm{TM})
} \\
&\qquad \times
\frac{
    P(f_\mathrm{TM})
}{
    A_\mathrm{TM}(f_\mathrm{TM})
} \, .
\end{aligned}
\end{equation}
Here, $H^\mathrm{model}_\mathrm{PCLG}(f)$ denotes the modeled PCLG.
The sensing-related quantity is computed from the photon-calibrator line at $f_\mathrm{pcal2}$ as
\begin{equation}
\label{eq:x_pcal2}
X(f_\mathrm{pcal2})
=
\kappa_\mathrm{TM} A_\mathrm{TM}(f_\mathrm{pcal2})
+
A_\mathrm{IM}(f_\mathrm{pcal2}) .
\end{equation}

\begin{equation}
\label{eq:s_sen}
\begin{aligned}
S(f_\mathrm{pcal2})
&=
\frac{1}{C_\mathrm{res}(f_\mathrm{pcal2})}
\Biggl(
    \left.\frac{d_\mathrm{err}}{d_\mathrm{pcal}}\right|_{f_\mathrm{pcal2}} \\
    &- D(f_\mathrm{pcal2})\,X(f_\mathrm{pcal2})
\Biggr)^{-1}.
\end{aligned}
\end{equation}
where
\begin{equation}
    \label{eq:s_res}
    C_\mathrm{res}(f)
    =
    H_C F_C(f) e^{-2\pi i f \tau_C} .
\end{equation}
In Eq.~\eqref{eq:s_sen}, the MN-stage contribution is neglected, since above $\sim 30$~Hz it is at the level of $\sim 0.1\%$ or less relative to the TM-stage contribution in the frequency range relevant to this analysis.
The optical gain correction factor $\kappa_C$ and the cavity pole frequency $f_\mathrm{cp}$ are then obtained as
\begin{align}
    \label{eq:s_tdcfs}
    \kappa_C
    &=
    \frac{|S(f_\mathrm{pcal2})|^2}{\Re[S(f_\mathrm{pcal2})]}, \\
    f_\mathrm{cp}
    &=
    - \frac{\Re[S(f_\mathrm{pcal2})]}{\Im[S(f_\mathrm{pcal2})]} \, f_\mathrm{pcal2}.
\end{align}

\bibliographystyle{apsrev4-2}
\bibliography{references}

\end{document}